\documentclass[aps,preprintnumbers,showpacs,superscriptaddress,nofootinbib,twocolumn]{revtex4}
\usepackage{epsfig,graphics}
\usepackage{amsfonts,amssymb,amsmath}            
\usepackage{amsthm}

\newcommand{\id}{\openone}

\newcommand{\comment}[1]{}
\newcommand{\ket}[1]{|#1\rangle}
\newcommand{\bra}[1]{\langle #1|}

\newcommand{\proj}[1]{|#1\rangle\!\langle #1|}
\newcommand*{\tr}{\mathrm{tr}}

\newcommand*{\cE}{\mathcal{E}}
\newcommand*{\cH}{\mathcal{H}}

\newcommand*{\cS}{\mathcal{S}}

\newcommand*{\LLambda}{\Lambda} 
\newcommand*{\llambda}{\lambda}

\theoremstyle{plain}

\theoremstyle{definition}

\newcommand{\cancel}[1]{}

\begin{document}

\title{Is a system's wave function in one-to-one correspondence with
  its elements of reality?}

\date{$12^{\text{th}}$ April, 2012}

\author{Roger Colbeck}
\affiliation{Perimeter Institute for Theoretical Physics, 31 Caroline Street
  North, Waterloo, ON N2L 2Y5, Canada}
\author{Renato Renner}
\affiliation{Institute for Theoretical Physics, ETH Zurich, 8093
Zurich, Switzerland}

\begin{abstract}
  Although quantum mechanics is one of our most successful physical
  theories, there has been a long-standing debate about the
  interpretation of the wave function---the central object of the
  theory. Two prominent views are that (i)~it corresponds to an
  element of reality, i.e.\ an objective attribute that exists before
  measurement, and (ii)~it is a subjective state of knowledge about
  some underlying reality.  A recent result [Pusey \emph{et al.}\
  arXiv:1111.3328] has placed the subjective interpretation into
  doubt, showing that it would contradict certain physically plausible
  assumptions, in particular that multiple systems can be prepared
  such that their elements of reality are uncorrelated.  Here we show,
  based only on the assumption that measurement settings can be chosen
  freely, that a system's wave function is in one-to-one
  correspondence with its elements of reality. This also eliminates
  the possibility that it can be interpreted subjectively.
\end{abstract}


\maketitle

\emph{Introduction.|}Given the wave function associated with a
physical system, quantum theory allows us to compute predictions for
the outcomes of any measurement.  Since a wave function corresponds to
an extremal state and is therefore maximally informative, one possible
view is that it can be considered an \emph{element of reality} of the
system, i.e., an objective attribute that exists before measurement.
However, an alternative view, often motivated by the probabilistic
nature of quantum predictions, is that the wave function represents
incomplete (subjective) knowledge about some underlying reality.
Which view one adopts affects how one thinks about the theory at a
fundamental level.

\begin{figure*}
\includegraphics[width=0.95\textwidth]{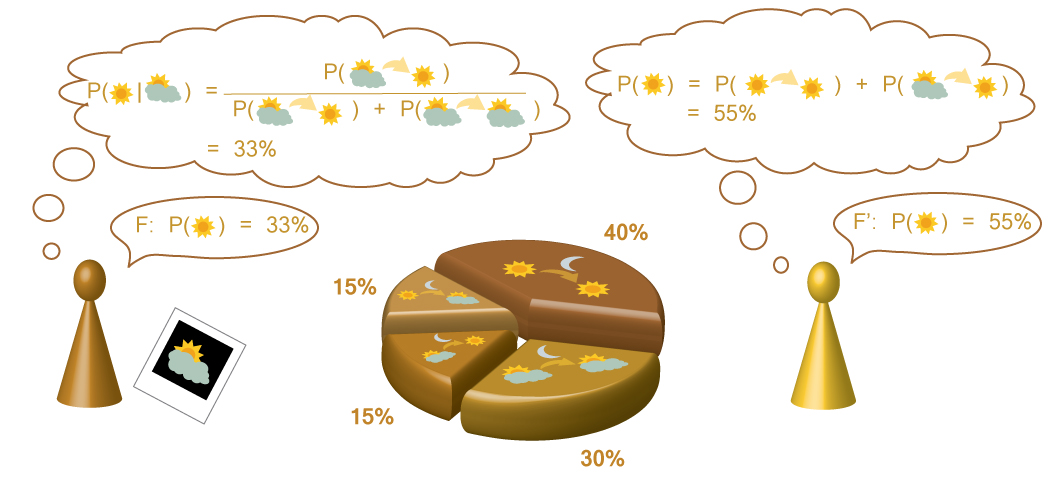}
\caption{\label{fig:weather}\textbf{Simple example illustrating the
    ideas.} Two meteorologists attempt to predict tomorrow's weather
  (whether it will be sunny or cloudy in a particular location).  Both
  have access to historical data giving the joint distribution of the
  weather on successive days. However, only the meteorologist on the
  left has access to today's weather, and consequently the two make
  different probabilistic forecasts, $F$ and $F'$.  Assuming that the
  processes relevant to the weather are accurately described by
  classical mechanics and thus deterministic, the list of elements of
  reality, $\LLambda$, may include tomorrow's weather, $X$. Such a
  list $\LLambda$ would then necessarily satisfy $\Gamma
  \leftrightarrow \LLambda \leftrightarrow X$ (for any arbitrary
  $\Gamma$) and therefore be complete (cf.\
  Eq.~\ref{eq_completelist}). However, the analogue of
  Eq.~\ref{eq_QMcompleteness}, $\LLambda \leftrightarrow F
  \leftrightarrow X$ would imply $X \leftrightarrow F \leftrightarrow
  X$. This Markov chain cannot hold for the non-deterministic
  forecasts $F$ and $F'$, which are hence not complete.  This is
  unlike the quantum-mechanical wave function, which gives a complete
  description for the prediction of measurement outcomes.  Note that
  this difference explains why, in contrast to the quantum-mechanical
  wave function, $F$ and $F'$ need not be included in $\Lambda$ and
  can therefore be considered subjective.}
\end{figure*}

To illuminate the difference between the above views, we give an
illustrative example.  Consider a meteorologist who gives a prediction
about tomorrow's weather (for example that it will be sunny with
probability $33 \%$, and cloudy with probability $67 \%$; see left
hand side of Fig.~\ref{fig:weather}).  We may assume that classical
mechanics accurately describes the relevant processes, so that the
weather depends deterministically on the initial conditions. The fact
that the prediction is probabilistic then solely reflects a lack of
knowledge on the part of the meteorologist on these conditions.  In
particular, the forecast is not an element of reality associated with
the atmosphere, but rather reflects the subjective knowledge of the
forecaster; a second meteorologist with different knowledge (see right
hand side of Fig.~\ref{fig:weather}) may issue an alternative forecast.

Moving to quantum mechanics, one may ask whether the wave function
$\Psi$ that we assign to a quantum system should be seen as a
subjective object (analogous to the weather forecast) representing the
knowledge an experimenter has about the system, or whether $\Psi$ is
an element of reality of the system (analogous to the weather being
sunny). This question has been the subject of a long debate, which
goes back to the early days of quantum theory~\cite{BV}.

The debate originated from the fact that quantum theory is inherently
probabilistic: even with a full description of a system's wave
function, the theory does not allow us to predict the outcomes of
future measurements with certainty.  This fact is often used to
motivate subjective interpretations of quantum theory, such as the
Copenhagen interpretation~\cite{Born26,Bohr28,Heisenberg30}, according
to which wave functions are mere mathematical objects that allow us to
calculate probabilities of future events.

Einstein, Podolsky and Rosen (EPR) advocated the view that the wave
function does not provide a complete physical description of
reality~\cite{EPR}, and that a higher, complete theory is possible. In
such a complete theory, any element of reality must have a counterpart
in the theory. Were quantum theory not complete, it could be that the
higher theory has additional parameters that complement the wave
function. The wave function could then be objective, i.e., uniquely
determined by the elements of reality of the higher
theory. Alternatively, the wave function could take the role of a
state of knowledge about the underlying parameters of the higher
theory. In this case, the wave function would not be uniquely
determined by these parameters and would therefore admit a
subjective interpretation. To connect to some terminology in the
literature (see for example Ref.~\citenum{HarSpek}), in the first case
the underlying model would be called \emph{$\psi$-ontic}, and in the
second case \emph{$\psi$-epistemic}. For some recent work in support
of a $\psi$-epistemic view, see for example
Refs.~\citenum{CFS,Spekkens_toy,LeiSpek}.

In some famous works from the 1960s, several constraints were placed
on higher descriptions given in terms of hidden
variables~\cite{KS,Bell_KS,Bell}, and further constraints have since
been highlighted~\cite{Hardy_ontbag,Montina,ChenMontina}.  In addition,
we have recently shown~\cite{CR_ext} that, under the assumption of free
choice, if quantum theory is correct then it is \emph{non-extendible},
in the sense of being maximally informative about measurement
outcomes.

Very recently, Pusey, Barrett and Rudolph~\cite{PBR} have presented an
argument showing that a subjective interpretation of the wave function
would violate certain plausible assumptions.  Specifically, their
argument refers to a model where each physical system possesses an
individual set of (possibly hidden) elements of reality, which are the
only quantities relevant for predicting the outcomes of later
measurements. One of their assumptions then demands that it is
possible to prepare multiple systems such that these sets are
statistically independent.

Here, we present a totally different argument to show that the wave
function of a quantum system is fully determined by its elements of
reality.
In fact, this implies that the wave function is in one-to-one
correspondence with these elements of reality (see the
Conclusions) and may therefore itself be considered an element of
reality of the system. These claims are derived under minimal assumptions,
namely that the statistical predictions of existing quantum theory are
correct, and that measurement settings can (in principle) be chosen
freely. In terms of the language of Ref.~\citenum{HarSpek}, this means
that any model of reality consistent with quantum theory and with free
choice is $\psi$-complete.

\smallskip

\begin{figure}
\includegraphics[width=0.5\textwidth]{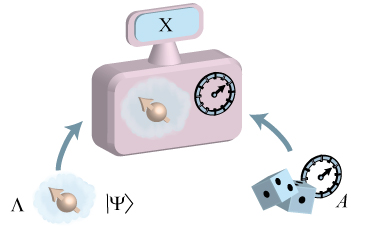}
\caption{\label{fig:setting}\textbf{Illustration of the setup.}  A
  system is prepared in a particular quantum state (specified by a
  wave function $\Psi$). The elements of reality, $\LLambda$, may
  depend on this preparation.  A measurement setting, $A$, is then
  randomly chosen, and the system is measured, producing an outcome,
  $X$.  We assume that $\LLambda$ is \emph{complete for the
    description of the system}, in the sense that there does not exist
  any other parameter that provides additional information (beyond
  that contained in $\LLambda$) about the outcome of any chosen
  measurement. In particular, $\Psi$ cannot provide more information
  than $\LLambda$.  Conversely, the non-extendibility of quantum
  theory~\cite{CR_ext} implies that $\LLambda$ cannot provide more
  information (about the outcome) than $\Psi$. Taken together, these
  statements imply that $\Psi$ and $\LLambda$ are informationally
  equivalent. From this and the fact that different quantum states
  generally lead to different measurement statistics, we conclude that
  $\Psi$ must be included in the list $\LLambda$ and is therefore an
  element of reality of the system.}
\end{figure}

\emph{General Model.|}In order to state our result, we consider a
general experiment where a system $\cS$ is prepared in a state
specified by a \emph{wave function} $\Psi$ (see
Fig.~\ref{fig:setting}). Then an experimenter chooses a
\emph{measurement setting} $A$ (specified by an observable or a family
of projectors) and records the \emph{measurement outcome}, denoted
$X$. Mathematically, we model $\Psi$ as a random variable over the set
of wave functions, $A$ as a random variable over the set of
observables, and $X$ as a random variable over the set of possible
measurement outcomes. Finally, we introduce a collection of random
variables, denoted $\Gamma$, which are intended to model all
information that is (in principle) available before the measurement
setting, $A$, is chosen and the measurement is carried
out. Technically, we only require that $\Gamma$ includes the wave
function $\Psi$.  In the following, when we refer to a \emph{list of
  elements of reality}, we simply mean a subset $\LLambda$ of
$\Gamma$. Furthermore, we say that $\LLambda$ is \emph{complete for
  the description of the system $\cS$} if any possible prediction
about the outcome $X$ of a measurement $A$ on $\cS$ can be obtained
from $\LLambda$, i.e., we demand that the Markov condition
\begin{align} \label{eq_completelist} \Gamma \leftrightarrow
  (\LLambda,A) \leftrightarrow X
\end{align}
holds~\footnote{$U \leftrightarrow V \leftrightarrow W$
  is called a \emph{Markov chain} if $P_{U|V=v} = P_{U|V=v, W=w}$ or,
  equivalently, if $P_{W|V=v} = P_{W|V=v,U=u}$, for all $u,v,w$ with
  strictly positive joint probability. Here and in the following, we
  use upper case letters for random variables, and lower case letters
  for specific values they can take.}.  Note that, using this
definition, the aforementioned result on the non-extendibility of
quantum theory~\cite{CR_ext} can be phrased as: \emph{The wave
  function $\Psi$ associated with a system $\cS$ is complete for the
  description of $\cS$}.

We are now ready to formulate our main technical claim.

\smallskip

\emph{Theorem.|}Any list of elements of reality, $\LLambda$, that is
complete for the description of a system $\cS$ includes the
quantum-mechanical wave function $\Psi$ associated with $\cS$ (in the
sense that $\Psi$ is uniquely determined by $\LLambda$).

\smallskip

\emph{Assumptions.|} The above claim is derived under the following
two assumptions, which are usually implicit in the literature. (We
note that very similar assumptions are also made in
Ref.~\citenum{PBR}, where, as already mentioned, an additional
statistical independence assumption is also used.)
\begin{itemize}
\item \emph{Correctness of quantum theory:} Quantum theory gives the
  correct statistical predictions. For example, the distribution of
  $X$ satisfies $P_{X|\Psi=\psi, A=a}(x) = \bra{\psi} \Pi^a_x
  \ket{\psi}$, where $\Pi^a_x$ denotes the projector corresponding to
  outcome $X=x$ of the measurement specified by $A = a$.

\item \emph{Freedom of choice:} Measurement settings can be chosen to
  be independent of any pre-existing value (in any
  frame)~\footnote{This assumption, while often implicit, is for
    instance discussed (and used) in Bell's work. In
    Ref.~\citenum{Bell_free}, he writes that ``the settings of
    instruments are in some sense free variables $\ldots$ [which]
    means that the values of such variables have implications only in
    their future light cones.''  This leads directly to the freedom of
    choice assumption as formulated here. We refer to the Supplemental
    Material for a more detailed discussion.}.  In particular, this
  implies that the setting $A$ can be chosen independently of
  $\Gamma$, i.e., $P_{A|\Gamma=\gamma} = P_{A}$~\footnote{In
    Ref.~\citenum{PBR}, this assumption corresponds to the requirement
    that a quantum system can be freely prepared according to one of a
    number of predefined states.}.
\end{itemize}
We note that the proof of our result relies on an argument presented
in Ref.~\citenum{CR_ext}, where these assumptions are also used
(see the Supplemental Material for more details).

\smallskip

\emph{Proof of the main claim.|} As shown in Ref.~\citenum{CR_ext},
under the above assumptions, $\Psi$ is complete for the description of
$\cS$. Since $\LLambda$ is included in $\Gamma$, we have in particular
\begin{align} \label{eq_QMcompleteness} \LLambda \leftrightarrow
  (\Psi,A) \leftrightarrow X \ .
\end{align}
Our argument then proceeds as follows. The above condition is
equivalent to the requirement that
\begin{align*}
  P_{X|\LLambda=\llambda,\Psi = \psi,A=a} = P_{X|\Psi=\psi,A=a}
\end{align*}
holds for all $\llambda, \psi, a$ that have a positive joint
probability, i.e., $P_{\LLambda \Psi A}(\llambda, \psi, a) > 0$.
Furthermore, because of the assumption that $\LLambda$ is a complete
list of elements of reality, Eq.~\ref{eq_completelist}, and because
$\Psi$ is by definition included in $\Gamma$, we have
\begin{align*}
  P_{X|\LLambda=\llambda, \Psi=\psi, A=a} = P_{X|\LLambda=\llambda, A=a} \ .
\end{align*}
Combining these expressions gives
\begin{align}  \label{eq_consistency}
  P_{X|\Psi = \psi,A=a} = P_{X|\LLambda=\llambda,A=a} \ ,
\end{align}
for all values $\llambda,\psi,a$ with $P_{\LLambda \Psi
  A}(\llambda,\psi, a)>0$.  Note that, using the free choice
assumption, we have $P_{\LLambda \Psi A} = P_{\LLambda \Psi} \times
P_A$, hence this condition is equivalent to demanding $P_{\LLambda
  \Psi}(\llambda, \psi)>0$ and $P_A(a)>0$.

Now consider some fixed $\LLambda=\llambda$ and suppose that there
exist two states, $\psi_0$ and $\psi_1$, such that $P_{\LLambda
  \Psi}(\llambda,\psi_0)>0$ and $P_{ \LLambda
  \Psi}(\llambda,\psi_1)>0$.  From Eq.~\ref{eq_consistency}, this
implies $P_{X|\Psi = \psi_0,A=a} = P_{X|\Psi = \psi_1,A=a}$ for all
$a$ such that $P_A(a)>0$. However, within quantum theory, it is easy
to choose the set of measurements for which $P_A(a)>0$ such that this
can only be satisfied if $\psi_0=\psi_1$.  This holds, for example, if
the set of measurements is tomographically complete.  Thus, for each
$\LLambda=\llambda$, there exists only one possible value of
$\Psi=\psi$ such that $P_{\Lambda \Psi}(\lambda, \psi) > 0$, i.e.,
$\Psi$ is uniquely determined by $\LLambda$, which is what we set out
to prove.

\smallskip

\emph{Discussion and Conclusions.|}We have shown that the quantum wave
function can be taken to be an element of reality of a system based on
two assumptions, the correctness of quantum theory and the freedom of
choice for measurement settings.  Both of these assumptions are in
principle experimentally falsifiable (see the Supplemental Material
for a discussion of possible experiments).

The correctness of quantum theory is a natural assumption given that
we are asking whether the quantum wave function is an element of
reality of a system.  Furthermore, a free choice assumption is
necessary to show that the answer is yes.  Without free choice, $A$
would be pre-determined and the complete list of elements of reality,
$\LLambda$, could be chosen to consist of the single element $X$. In
this case, Eq.~\ref{eq_completelist} would be trivially satisfied.
Nevertheless, since the list $\LLambda = \{X\}$ does not uniquely
determine the wave function, $\Psi$, we could not consider $\Psi$ to
be an element of reality of the system. This shows that the wave
function would admit a subjective interpretation if the free choice
assumption was dropped.

We conclude by noting that, given any complete list of elements of
reality, $\LLambda$, the non-extendibility of quantum theory,
Eq.~\ref{eq_QMcompleteness}, asserts that any information contained in
$\LLambda$ that may be relevant for predicting measurement outcomes
$X$ is already contained in the wave function $\Psi$. Conversely, the
result shown here is that $\Psi$ is included in $\LLambda$.  Since
these are two seemingly opposite statements, it is somewhat intriguing
that the second can be inferred from the first, as shown in this
Letter. Furthermore, taken together, the two statements imply that
$\Psi$ is in one-to-one correlation to $\LLambda$.  This sheds new
light on a question dating back to the early days of quantum
theory~\cite{EinsteinSchroedinger}, asking whether the wave function
is in one-to-one correlation with physical reality.  Interpreting
$\LLambda$ as the state of physical reality (or the ontic state), our
result asserts that, under the free choice assumption, the answer to
this question is yes.

\smallskip

\noindent\emph{Acknowledgements.|}We thank Christian Speicher and
Oscar Dahlsten for posing questions regarding Ref.~\citenum{PBR},
which initiated this work, and Jonathan Barrett for explaining their
result. We also thank Matthias Christandl, Alberto Montina, L\'idia
del Rio and Robert Spekkens for discussions and L\'idia del Rio for
illustrations. This work was supported by the Swiss National Science
Foundation (through the National Centre of Competence in Research
``Quantum Science and Technology'' and grant No.\ 200020-135048) as
well as the European Research Council (grant No.\ 258932).  Research
at Perimeter Institute is supported by the Government of Canada
through Industry Canada and by the Province of Ontario through the
Ministry of Research and Innovation.

\section*{SUPPLEMENTAL MATERIAL}

\section{Additional discussion of the assumptions}

Our work relies on two assumptions, which we discuss in separate
subsections.  We note that these assumptions are essentially those
already used in~\cite{CR_ext}, upon which this work builds.

For the following exposition, it is convenient to introduce the
concept of \emph{spacetime variables
  (SVs)}~\cite{CR_ext}. Mathematically, these are simply random
variables (which take values from an arbitrary set) together with
associated coordinates $(t,r_1,r_2,r_3) \in \mathbb{R}^4$. A SV may be
interpreted physically as a value that is accessible at the spacetime
point specified by these coordinates (with respect to a given
reference frame).

The variables described in the main text, $A$, $X$, and $\Gamma$, can
readily be modelled as SVs. In particular, the coordinates of $A$
should specify the spacetime point where the measurement setting (for
measuring the system $\cS$) is chosen. Accordingly, the coordinates of
$X$ correspond to an (arbitrary) point in the spacetime region where
the measurement outcome is available.  We therefore assign coordinates
such that $X$ is in the future lightcone of $A$, whereas no SV in the
set $\Gamma$ (which models any information available before the
measurement) should lie in the future lightcone of $A$.

\subsection{Correctness of quantum theory}

This assumption refers to the statistical predictions about
measurement outcomes that can be made within standard quantum theory
(i.e., it does not make reference to any additional parameters of a
potential higher theory\footnote{In particular, in a higher theory
  that has additional hidden parameters, the predictions of quantum
  theory should be recovered if these parameters are ignored, which
  corresponds mathematically to averaging over them.}).  Following the
treatment in~\cite{CR_ext}, we subdivide the assumption into two
parts.

\begin{itemize}
\item \emph{QMa:} Consider a system whose state is described by a wave
  function, $\Psi$, corresponding to an element of a Hilbert space
  $\cH_S$. According to quantum theory, any measurement setting $A=a$
  is specified by a family of projectors, $\{\Pi^a_x\}_{x}$,
  parameterized by the possible outcomes $x$, such that $\sum_x
  \Pi^a_x = \id_{\cH_S}$.\footnote{Naimark's dilation theorem asserts
    that any measurement specified by a Positive Operator Valued
    Measure can be seen as a projective measurement on a larger
    Hilbert space.} The assumption demands that the probability of
  obtaining output $X=x$ when the system is measured
  with 
  setting $A=a$ is given by
  \begin{align*}
  P_{X|A=a}(x) = \tr(\proj{\Psi} \Pi^a_x) \ .
  \end{align*}
 
\item \emph{QMb:} Consider again a measurement as in
  Assumption~\emph{QMa} described by projectors $\{\Pi^a_x\}_x$ on
  $\cH_S$. According to quantum theory, for any fixed choice of the
  setting $A = a_0$, the measurement can be modelled as an isometry
  $\cE_{S \to SE}$ from states on $\cH_S$ to states on a larger system
  $\cH_S \otimes \cH_E$ (involving parts of the measurement apparatus
  and the environment) such that the restriction of $\cE_{S \to SE}$
  to the original system corresponds to the initial measurement, i.e.,
  formally
  \begin{align*}
    \tr_E \cE_{S \to SE}(\proj{\Psi})  = \sum_x \Pi^{a_0}_x \proj{\Psi} \Pi^{a_0}_x \ ,
  \end{align*}
  where $\tr_E$ denotes the partial trace over
  $\cH_E$~\cite{Stinespring}. The assumption then demands that for all
  $A=a$ and for any measurement (defined by a family of projectors,
  $\{\Pi^b_y\}_{y}$, parameterized by the possible outcomes $y$)
  carried out on system $E$, the joint statistics of the outcomes are
  given by
  \begin{align*}
    P_{X Y | A=a, B=b}(x, y) = \tr\bigl(\cE_{S \to SE}(\proj{\Psi}) \Pi^a_x
    \otimes \Pi^b_y\bigr) \ .
  \end{align*}
\end{itemize}

Note that both assumptions refer to the Born rule~\cite{Born26} for
the probability distribution of measurement outcomes. In
Assumption~\emph{QMa}, the rule is applied to a measurement on a
single system, whereas Assumption~\emph{QMb} demands that the rule
also applies to the joint probability distribution involving the
outcome of (arbitrary) additional measurements.


\subsection{Freedom of choice}
The assumption that measurement settings can be chosen freely is often
left implicit in the literature. This is also true, for example, for
large parts of Bell's work, although he later mentioned the
assumption explicitly~\cite{Bell_free}.

The notion of freedom of choice can be expressed mathematically using
the language of SVs. We say that a SV $A$ is \emph{free with respect
  to a set of SVs $\Omega$} if
\begin{align*}
  P_{A\Omega'} = P_{A} \times P_{\Omega'}
\end{align*}
holds, where $\Omega'$ is the set of all SVs from $\Omega$ whose
coordinates lie outside the future lightcone of $A$. This captures the
idea that $A$ should be independent of any ``pre-existing'' values
(with respect to any reference frame).

This definition is motivated by the following notion of causality.
For two SVs $A$ and $B$, we say that $B$ \emph{could have been caused
  by} $A$ if and only if $B$ lies in the future lightcone of $A$.
Within a relativistic spacetime structure, this is equivalent to
requiring the time coordinate of $B$ to be larger than that of $A$ in
all reference frames.  Using this notion of causality, our definition
that $A$ is free with respect to $\Omega$ is equivalent to demanding
that all SVs in $\Omega$ that are correlated to $A$ could have been
caused by $A$.

Connecting to the main text, we note that (by definition) all SVs in
the set $\Gamma$ defined there lie outside the future lightcone of the
spacetime point where the measurement setting $A$ is chosen. The
requirement for $A$ to be free with respect to $\Gamma$ thus simply
reads $P_{A\Gamma} = P_A \times P_{\Gamma}$.

In Ref.~\citenum{CR_ext} (upon which the present result is based), the
free choice assumption is used in a more general bipartite scenario.
There, two measurements are carried out at spacelike separation, one
of which has setting $A$ and outcome $X$ and the other has setting $B$
and outcome $Y$. In addition, as in our main argument, we consider
arbitrary additional (pre-existing) information $\Gamma$.  The
assumption that $A$ and $B$ are chosen freely (i.e., such that they
are uncorrelated with any variables in their past in any frame) then
corresponds mathematically to the requirements $P_{A|BY\Gamma} = P_A$
and $P_{B|AX\Gamma} = P_B$.  We remark that these conditions are not
obeyed in the de Broglie-Bohm model~\cite{deBroglie,Bohm} if one
includes the wave function as well as the hidden particle trajectories
in $\Gamma$.

It is also worth making a few additional remarks about the connection
to other work.  As mentioned in the main text, in
Ref.~\citenum{Bell_free}, Bell writes that ``the settings of
instruments are in some sense free variables $\ldots$ [which] means
that the values of such variables have implications only in their
future light cones.'' When formalized, this gives the above
definition.  However, in spite of the motivation given in the above
quote, the mathematical expression Bell writes down corresponds to a
weaker notion that only requires free choices to be independent of
pre-existing hidden parameters (but does not include pre-existing
measurement outcomes). This weaker requirement is (as he acknowledges)
a particular implication of the full freedom of choice assumption. We
imagine that the reason for Bell's reference to this weaker
implication is that it is sufficient for his purpose when combined
with another assumption, known as local causality. Indeed, the weaker
implication of free choice together with Bell's local causality are
also sufficient to prove our result.  Furthermore, in the literature
the weaker notion is sometimes taken to be the definition of free
choice, rather than an implication of it.

\section{Connection to experiment}
Our main argument is based on the assumption that measurement outcomes
obey the statistical predictions of quantum theory, and it is
interesting to consider how closely experimental observations come to
obeying these predictions.  For the argument in Ref.~\citenum{CR_ext},
which leads to Eq.~2 
in the main text, this assumption is divided into two parts, as
mentioned above.

The first part of the assumption has already been subject to
experimental investigation (see Refs.~\citenum{PBSRG,SSCRT}), giving
results compatible with quantum theory to within experimental
tolerance.  Note that, although no experimental result can establish
the Markov chain condition of Eq.~2 
precisely, the observed data can be used to bound how close (in trace
distance) the Markov chain condition is to holding (see
Ref.~\citenum{SSCRT} for more details).

The second part of the assumption has not seen much experimental
attention to date.  However, were we to ever discover a measurement
procedure that is demonstrably inconsistent with unitary dynamics on
the microscopic scale, this would falsify the assumption and point to
new physics.

The freedom of choice assumption is more difficult to probe
experimentally, since it is stated in terms of $\Gamma$, which is
information in a hypothetical higher theory.  Nevertheless, it would
be possible to falsify the assumption in specific cases, for example
using a device capable of predicting the measurement settings before
they were chosen.

\end{document}